\documentclass[10pt,aps,prl,superscriptaddress,twocolumn]{revtex4-1}

\usepackage[utf8]{inputenc}
\usepackage[english]{babel}
\usepackage{times}
\usepackage{color}
\usepackage{graphicx}
\usepackage{import}
\usepackage{amsmath}
\usepackage{braket}
\usepackage{url}
\usepackage{breakurl}
\usepackage[hyperindex,breaklinks]{hyperref}

\newcommand*\diff{\mathop{}\!\mathrm{d}}

\usepackage[normalem]{ulem}

\begin{document}
\title{Optimizing Schedules for Quantum Annealing}
\author{Daniel Herr}
\affiliation{Theoretische Physik, ETH Zurich, 8093 Zurich, Switzerland}
\affiliation{Quantum Condensed Matter Research Group, CEMS, RIKEN, Wako-shi 351-0198, Japan}
\author{Ethan Brown}
\affiliation{Theoretische Physik, ETH Zurich, 8093 Zurich, Switzerland}
\author{Bettina Heim}
\affiliation{Theoretische Physik, ETH Zurich, 8093 Zurich, Switzerland}
\author{Mario K\"onz}
\affiliation{Theoretische Physik, ETH Zurich, 8093 Zurich, Switzerland}
\author{Guglielmo Mazzola}
\affiliation{Theoretische Physik, ETH Zurich, 8093 Zurich, Switzerland}
\author{Matthias Troyer}
\affiliation{Theoretische Physik, ETH Zurich, 8093 Zurich, Switzerland}
\affiliation{Quantum Architectures and Computation Group, StationQ, Microsoft Research,
Redmond, WA 98052, USA}

\begin{abstract}
Classical and quantum annealing are two heuristic optimization methods that search for an optimal solution by slowly decreasing thermal or quantum fluctuations. Optimizing annealing schedules is important both for performance and fair comparisons between classical annealing, quantum annealing, and other algorithms. Here we present a heuristic approach for the optimization of annealing schedules for quantum annealing and apply it to 3D Ising spin glass problems. We find that if both classical and quantum annealing schedules are similarly optimized, classical annealing outperforms quantum annealing for these problems when considering the residual energy obtained in slow annealing. However, when performing many repetitions of fast annealing, simulated quantum annealing is seen to outperform classical annealing for our benchmark problems.
\end{abstract}
\maketitle

Optimization is a fundamental task underlying many important problems in a wide range of disciplines. Solving an optimization problem means  finding a configuration that (approximately) minimizes some cost function. This cost function can be interpreted as the Hamiltonian of a physical system and its minimization then corresponds to finding the ground state. In particular, discrete combinatorial optimization problems can be mapped~\cite{ising_rep,travelling_salesman} to Ising spin glass problems, given by the Hamiltonian
\begin{equation}
 H_p = - \sum_{i<j} J_{ij} s_i s_j - \sum_i h_i s_i.
 \label{eq:H_P}
\end{equation}
Here, the spins $s_i=\pm1$ are coupled pairwise through the $J_{ij}$ terms and individually to external magnetic fields $h_i$. Finding the ground state of such Ising spin glass problems is a nondeterministic-polynomially (NP) hard problem~\cite{barahona} for which no efficient, polynomial time algorithms exist. Given the importance of these problems in many application areas, finding efficient optimization algorithms is an important endeavor.

One powerful heuristic algorithm to find low energy configurations of such a problem is simulated classical annealing (CA)~\cite{kirkpatrick1983,Metropolis}. There, the system is first initialized to a random configuration at a high temperature and then gradually cooled during Monte Carlo simulations. This allows the system's configurations to escape from local minima and to thermally relax towards lower energy configurations~\cite{sa_and_sa_conv_cond}.

Quantum annealing (QA)~\cite{Ray1989,Finnila1994,Kadowaki1998,idea_of_qa,qareview} is a related method, which has gained attention due to its implementation in commercial devices~\cite{6802426,PhysRevB.82.024511,Johnson2011}. Instead of thermal fluctuations quantum fluctuations are used to drive the system out of local minima.
During QA the system evolves from a trivial ground state of a quantum mechanical driver Hamiltonian $\mathcal{H}_D$ to the solution of the problem Hamiltonian from Eq.~(\ref{eq:H_P}). The evolution of the  Hamiltonian is controlled by a  parameter $s(t) \in \left[0 ,1\right]$:
\begin{equation}
 \mathcal{H}=s \mathcal{H}_P + \left(1-s\right) \mathcal{H}_D.
 \label{eq:generalhamil}
\end{equation}
This enables a transition from  $\mathcal{H}_D$ to $\mathcal{H_P}$. For this the Ising variables are represented by quantum spin-1/2 variables, which results in
\begin{equation}
	\mathcal{H}_P = - \sum_{i<j} J_{ij} \sigma_z^i \sigma_z^j - \sum_i h_i \sigma_z^i,
\end{equation}
where $\sigma_z^j$ is the Pauli-$z$ matrices acting on spin $j$ 
The driver Hamiltonian is given by
\begin{equation}
\mathcal{H}_D = - \Gamma_0 \sum_{i} \sigma_x^i.
\end{equation}

Alternatively one could just vary the transverse field $\Gamma$ while $\mathcal{H}_P$ is kept fixed:
\begin{equation}
	\mathcal{H} = \mathcal{H}_P + \left(1-s\right) \mathcal{H}_D.
	\label{eq:spechamiltonian}
\end{equation}
We will use this schedule for our simulations.

We implement QA using a discrete time quantum Monte Carlo (QMC) simulation, because an exact simulation of QA is exponentially hard. We use the path integral Monte Carlo (PIMC) method to map a 3 dimensional quantum system into $3+1$ classical dimensions by the introduction of the imaginary time dimension~\cite{suzuki_orig}. It should be noted that this version of simulated quantum annealing (SQA), is indicative of the performance of a physical quantum annealer~\cite{mazzola2017quantum,Guglielmo,jiang2016scaling,google}.

Recent comparisons between quantum and classical annealing for two-dimensional (2D) Ising spin glass systems have not shown any quantum speedup~\cite{Troels,Heimb15}. The lack of speedup in this case may be due to the energy landscape of 2D spin glasses, with shallow but broad barriers that are easier to thermally surmount than to tunnel through~\cite{3dbetter}.

Extending such simulations to three dimensions (3D) we find that whether classical or quantum annealing is better depends on the specific choice of annealing schedules. An apparent advantage of one method may simply be due to a bad or suboptimal choice of annealing schedule for the other method. To achieve a fair comparison both classical and quantum annealing schedules thus need to be fairly optimized. In this Letter we thus introduce a heuristic optimization method for quantum annealing, generalizing heuristics used to to optimize classical annealing. We show that this heuristic leads to improvements over na\"ive schedules and allows fair comparisons of classical versus quantum annealing. 

\paragraph{Adaptive Schedule}
In classical annealing the control parameter is the temperature, whose change over time is given by an annealing schedule. If the change is constant we call it a linear schedule and the update rule for the temperature is given by
\begin{equation}
\beta(t)=\beta_0+\tilde{\lambda} t,
\end{equation}
or in a discretized form
\begin{equation}
 \beta_{k} = \beta_0+\lambda k = \beta_{k-1}  + \lambda,
\end{equation}
where $\beta_k$ is the inverse temperature at the $k$-th update sweep. For such a linear schedule, only the initial and final values of $\beta$ may need to be optimized, with intermediate values that are obtained by linear interpolation.

Instead of guessing a schedule, one can determine optimized adaptive schedules by using a heuristic algorithm to optimize the schedule~\cite{Huang86,sim_annealing,Fischer1995} in such a way that interesting regions where large changes to configurations may occur (e.g.\ close to phase transitions) are passed through more slowly.

An indicator for the size of a temperature step can be the specific heat
\begin{equation}
 C_V = k_B \beta^2 \frac{\diff^2}{\diff \beta^2} \log{Z} = - k_B \beta^2 \frac{\diff}{\diff \beta} \langle E \rangle,
 \label{eq:spec_heat}
\end{equation}
where  $Z$ is the partition function. $C_V$ can be calculated from the fluctuation-dissipation theorem as
\begin{equation}
 \beta^2C_V \equiv \sigma \equiv -\frac{\diff}{\diff\beta} \langle E \rangle = \langle {E}^2\rangle - {\langle  E \rangle}^2.
 \label{eq:cbefore}
\end{equation}

In order to achieve a constant decrease in energy at each step, one aims for this quantity to be constant throughout the annealing process:
\begin{equation}
 \frac{\diff \langle E \rangle}{\diff t} = -\lambda,
\end{equation}
where the scale factor $\lambda$ sets the targeted change in energy. The quantity $\langle E \rangle$ is only implicitly dependent on time through its dependence on the temperature, such that one can perform the chain rule
\begin{equation}
 \frac{\diff \langle E \rangle}{\diff \beta} \frac{\diff \beta}{\diff t} = -\lambda.
\end{equation}
Inserting Eq.~(\ref{eq:cbefore}) one obtains 
\begin{equation}
\frac{\diff \beta}{\diff t} \sigma\left(\beta\right) = \lambda,
\end{equation}
which can be discretized to obtain the an update rule for the adaptive schedule
\begin{equation}
 \beta_{k+1} = \beta_{k} + \frac{\lambda}{\sigma_k}.
\end{equation}
It can be observed that this schedule traverses slower in regions close to the phase transition, where $C_V$ is large. 

In order to derive a similarly optimized schedule for quantum annealing we use a quantity akin to the specific heat in classical systems. Starting from the quantum mechanical partition function
\begin{equation}
 Z(s) = \text{tr}\left(e^{-\beta \mathcal{H}(s)}\right)
\end{equation}
we first derive an adaptive schedule for Eq.~(\ref{eq:generalhamil}). Substituting the derivative with respect to $\beta$ by one with respect to the quantum control parameter $s$ in Eq.~(\ref{eq:spec_heat}) we define
\begin{equation}
C(s)=\frac{1}{\beta}\frac{\diff^2}{\diff s^2}\log\left(Z\right),
\end{equation}
which will determine the annealing schedule. Performing the derivatives we obtain
\begin{eqnarray}
 C(s)&=& - \frac{d}{ds} \langle \mathcal{H}_P - \mathcal{H}_D\rangle_s \nonumber \\
 &=& \beta \left(\langle \left( \mathcal{H}_P - \mathcal{H}_D\right)^2 \rangle_s - \langle \mathcal{H}_P - \mathcal{H}_D \rangle_s^2\right)
\end{eqnarray}
Note, that the expectation values in this equation are dependent on the parameter $s$.
Similar to the classical procedure, we aim for a constant change in time:
\begin{equation}
 \frac{\diff s}{\diff t} C\left(s\right) = \lambda.
\end{equation}
Thus we obtain an update rule for the quantum annealing schedule:
\begin{equation} 
 s_{k+1} = s_k + \frac{1}{\beta}\frac{\lambda}{\sqrt{\langle  {(\mathcal{H}_P - \mathcal{H}_D)}^2 \rangle_s - \langle  \mathcal{H}_P-\mathcal{H}_D \rangle_s^2}}
\end{equation}

For the alternative quantum annealing procedure of Eq.~(\ref{eq:spechamiltonian}), a similar derivation leads to a simpler rule
\begin{equation}
 s_{k+1} = s_k + \frac{1}{\beta \Gamma_0}\frac{\lambda}{\sqrt{1 - \langle \sigma_x\rangle_s^2 }}.
\end{equation}
Note that $\Gamma_0$ corresponds to the initial transverse field of the annealing schedule and an appropriate value needs to be determined before any efficient annealing run.

For classes of random instances one does not optimize the schedule for each individual instance but instead optimizes over the average of a limited set of instances, and use the average of this set to define an adaptive schedule for the entire class.

\paragraph{Results}

In the remainder of this work we will use the latter schedule, obtained by an ensemble average of $\langle \sigma_x \rangle_s$ over a set of $1000$ random, uniformly distributed 3D Ising spin glass instances on a simple cubic lattice. We investigate 3D spin glasses because of the argument~\cite{3dbetter} that random 3D Ising spin glasses are more likely to profit from quantum tunneling since they exhibit a non-zero temperature phase transition~\cite{3dphasetransition}.

\begin{figure}
 \includegraphics[width=\linewidth]{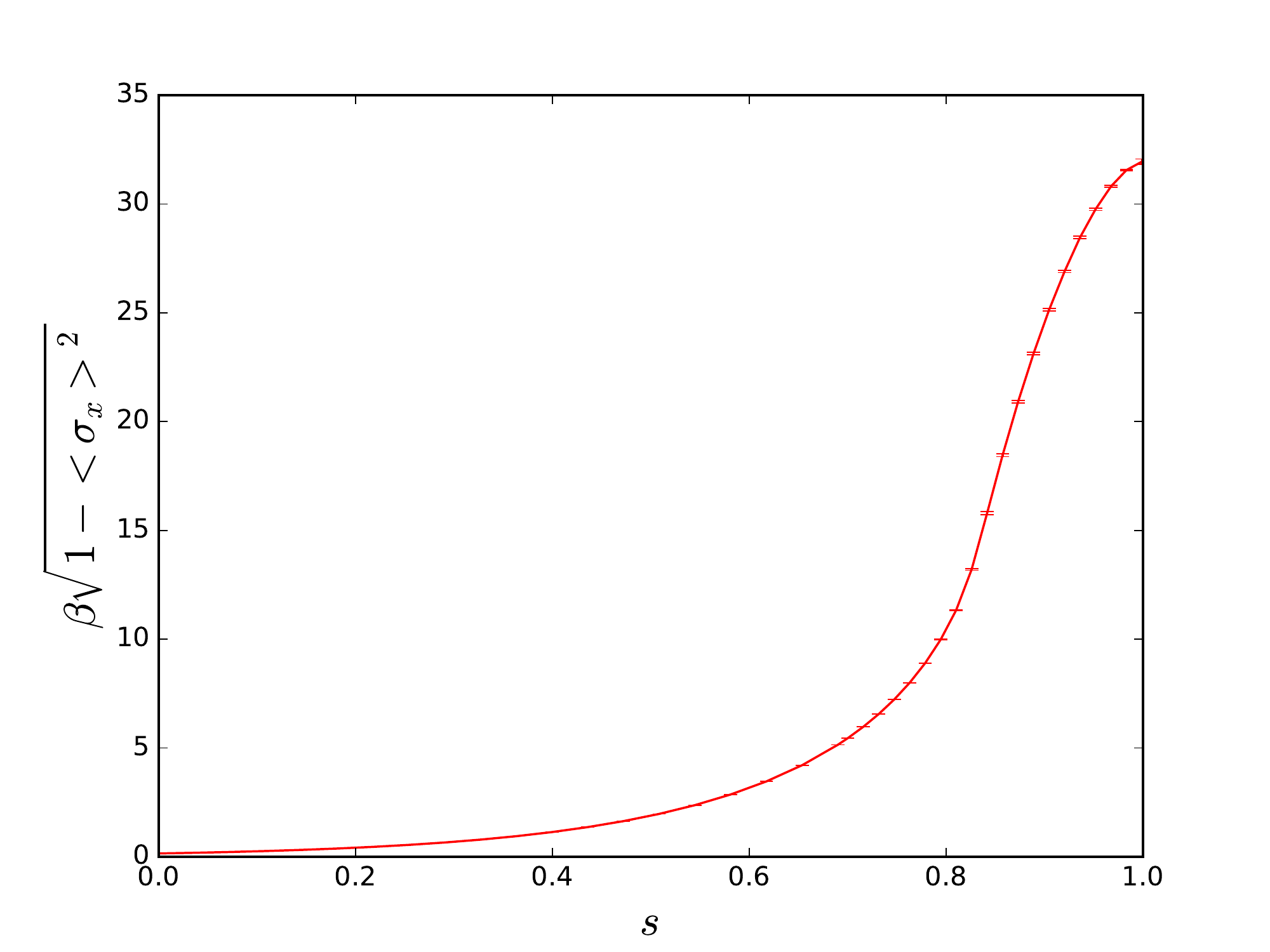}
 \caption{\label{fig:CQ} Expectation value of the step size denominator as a function of $s$. Thus it can be seen, that initially, for small values of $s$ annealing proceeds fast and slows down towards the end. Measurements were performed at $\beta = 32$ and $\Gamma_0 = 10$. The expectation value for this plot was obtained using a set of 100 random 3D Ising instances with couplings chosen uniformly from $\left[-1,1\right]$.
 The transverse field relates to the control parameter with $\Gamma(s) = \left(1-s \right)\Gamma_0$.}
\end{figure}

In Figure~\ref{fig:CQ} we we show the expectation value of the denominator of the step size, averaged over 1000 instances. In the beginning, when the transverse field is strong, the system is close to an eigenstate of $\sigma_x$ and the problem Hamiltonian does not influence the dynamics much. Thus, the step size is large.

In order to find an optimized schedule, the transverse field $\Gamma_0$ has to be optimized. This corresponds to an optimization of the starting value of the schedule and yielded $\Gamma_0=1.5$. The final value of the transverse field needs to be $\Gamma = 0$ in order to recover the problem Hamiltonian at the end of the annealing process.

\begin{figure}[t]
 \includegraphics[width=\linewidth]{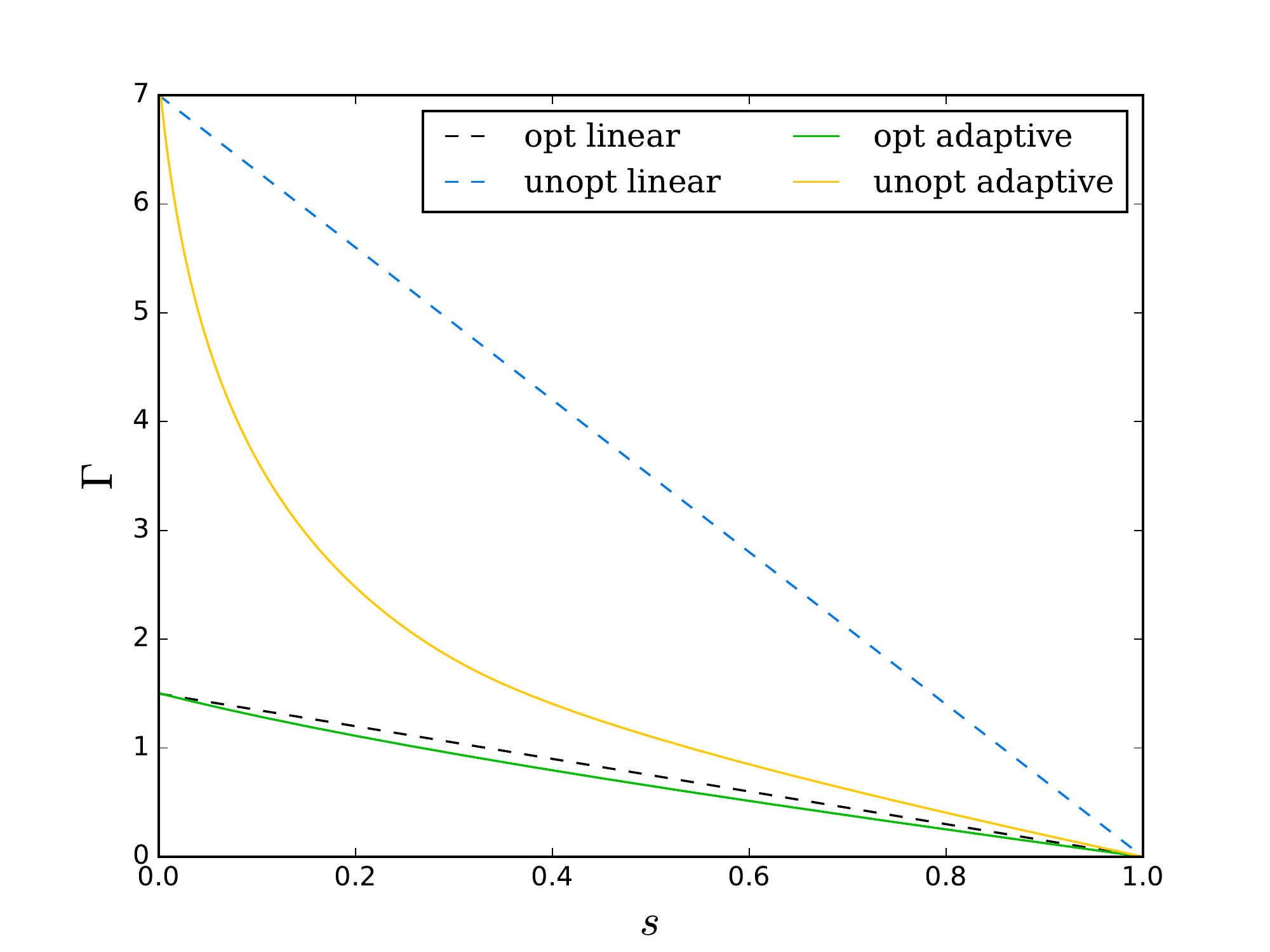}
 \caption{\label{fig:schedules}Plot of linear schedule and adaptive schedules with optimized and unoptimized initial transverse Field $\Gamma_0$. Close to the optimal values ($\Gamma_0 \approx 1.5$) one can see only minor differences in the two schedules, but for unoptimized starting values ($\Gamma_0 = 7$) the adaptive schedule anneals much faster at the beginning such that less time is spent in unimportant regions.}
\end{figure}

\begin{figure}[b]
 \includegraphics[width=\linewidth]{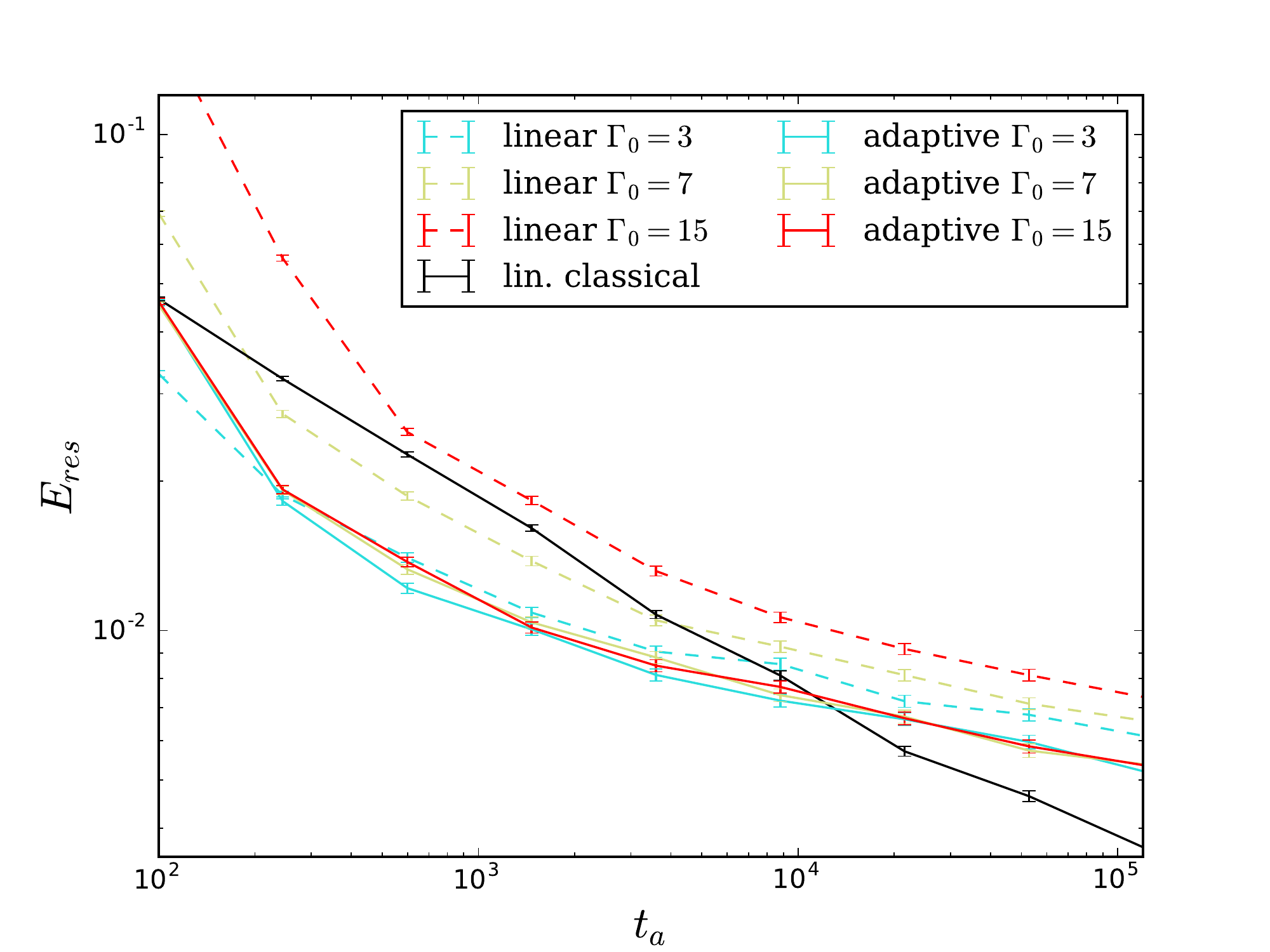}
 \caption{\label{fig:unopt}Comparison of median residual energy between linear and adaptive schedules at different starting values $\Gamma_0$. The linear schedule steadily degrades with increasingly suboptimal $\Gamma_0$ whereas the adaptive schedule  degrades only slightly.}
\end{figure}

\begin{figure}[t]
 \includegraphics[width=\linewidth]{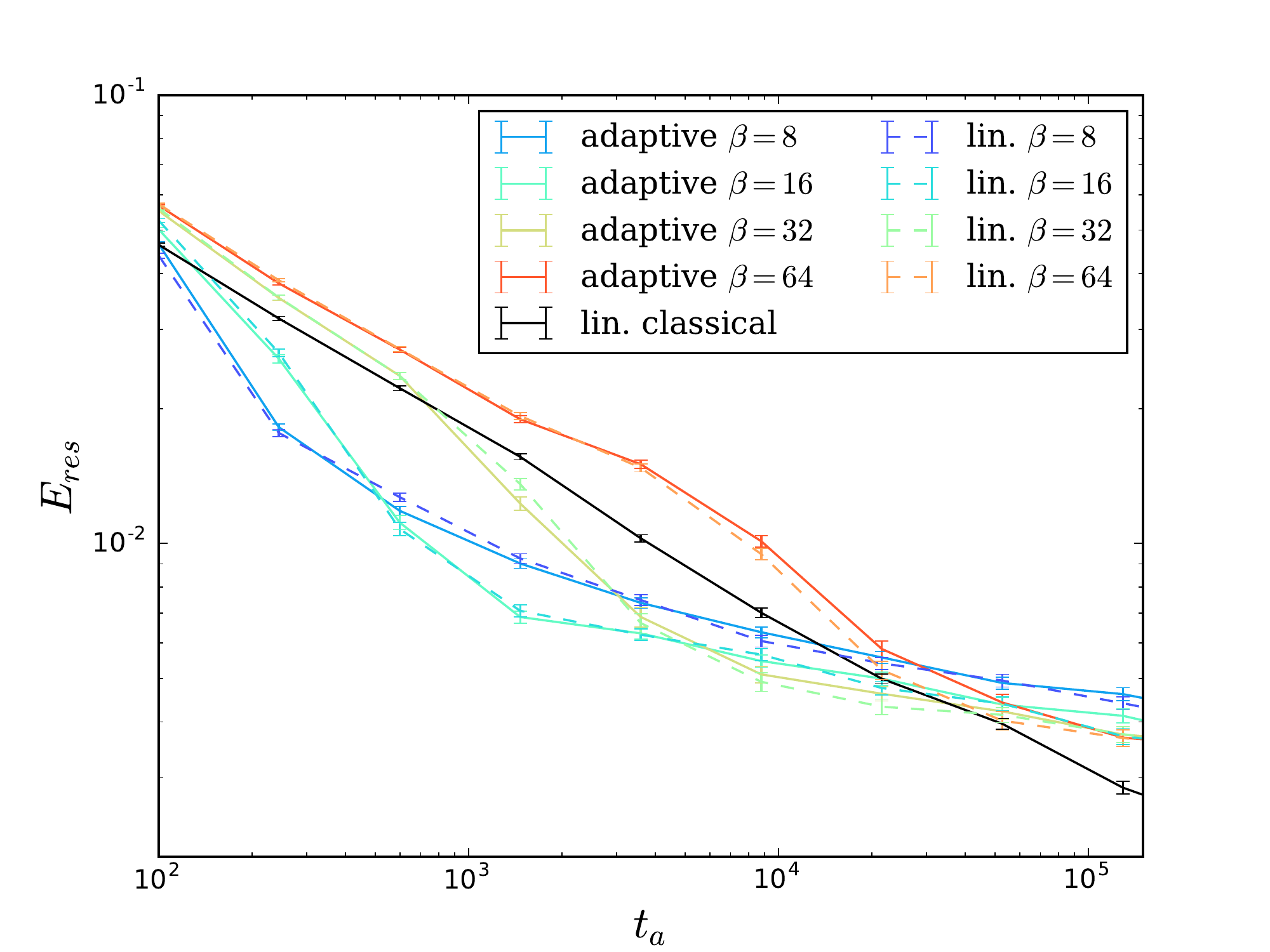}
 \caption{\label{fig:converged} Comparison of median residual energy for different values of $\beta$ between adaptive, linear QA schedules and linear CA. One can see almost no qualitative difference between the two schedules when only optimized parameters are used.}
\end{figure}

Figure~\ref{fig:schedules} shows the linear and the adaptive schedules for optimized and unoptimized initial field $\Gamma_0$. Consistent with Fig.~\ref{fig:CQ} one can see that the adaptive schedules are slower in regions, where the transverse field is weak. The residual energy of these schedules is shown Fig~\ref{fig:unopt}, which demonstrates improved performance of the adaptive schedules over the linear ones.  The improvement is especially large if one starts with unoptimized large initial transverse fields $\Gamma_0$. In those cases the adaptive schedule quickly reduces the transverse field and then slows down when entering the interesting parameter region. Conversely, one can see in Figure~\ref{fig:converged} that, for an optimized $\Gamma_0$ the optimized schedule is close to a linear one, with similar performance.

Since neither CA nor QA are guaranteed to find the exact ground state, we will first investigate the residual energy $E_{\rm res}= E - E_{0}$ as a metric to compare the efficiency of different annealing schemes. It is defined as the difference between the  energy of the (local) minimum $E$ found in an annealing run and the ground state energy $E_0$. The residual energy is a function of annealing time $t_a$, which in our simulations is given in units of Monte Carlo sweeps.
The behavior of SQA on 3D Ising spin glasses is similar to that of 2D Ising spin glasses~\cite{Heimb15}. SQA, running fast, can initially lower the energy much faster than CA but tends to get stuck in local minima.  Figure~\ref{fig:converged} shows that this initial fast convergence happens earlier with higher temperature. Yet, for every choice of inverse temperature, SQA suffers from getting stuck in local minima. CA does not display this problem, which is why CA outperforms any SQA run after sufficient annealing time.

As an alternative approach to slow annealing, many fast annealing runs might be beneficial for some problems~\cite{damian}. The total effort here is the product of the annealing time and the  number of repetitions $R$ required to find the ground state with a probability $\bar{s}$. Denoting the probability to find the ground state in a single execution by $p$ the required number of repetitions is  $ R={\log(1-p)}/{\log(1-\bar{s})}$ \cite{Troels}.
To compare the different annealing approaches we choose $t_a$ to optimize the total computational effort $t_a \cdot R$ for each system size.

\begin{figure}[t]
 \includegraphics[width=\linewidth]{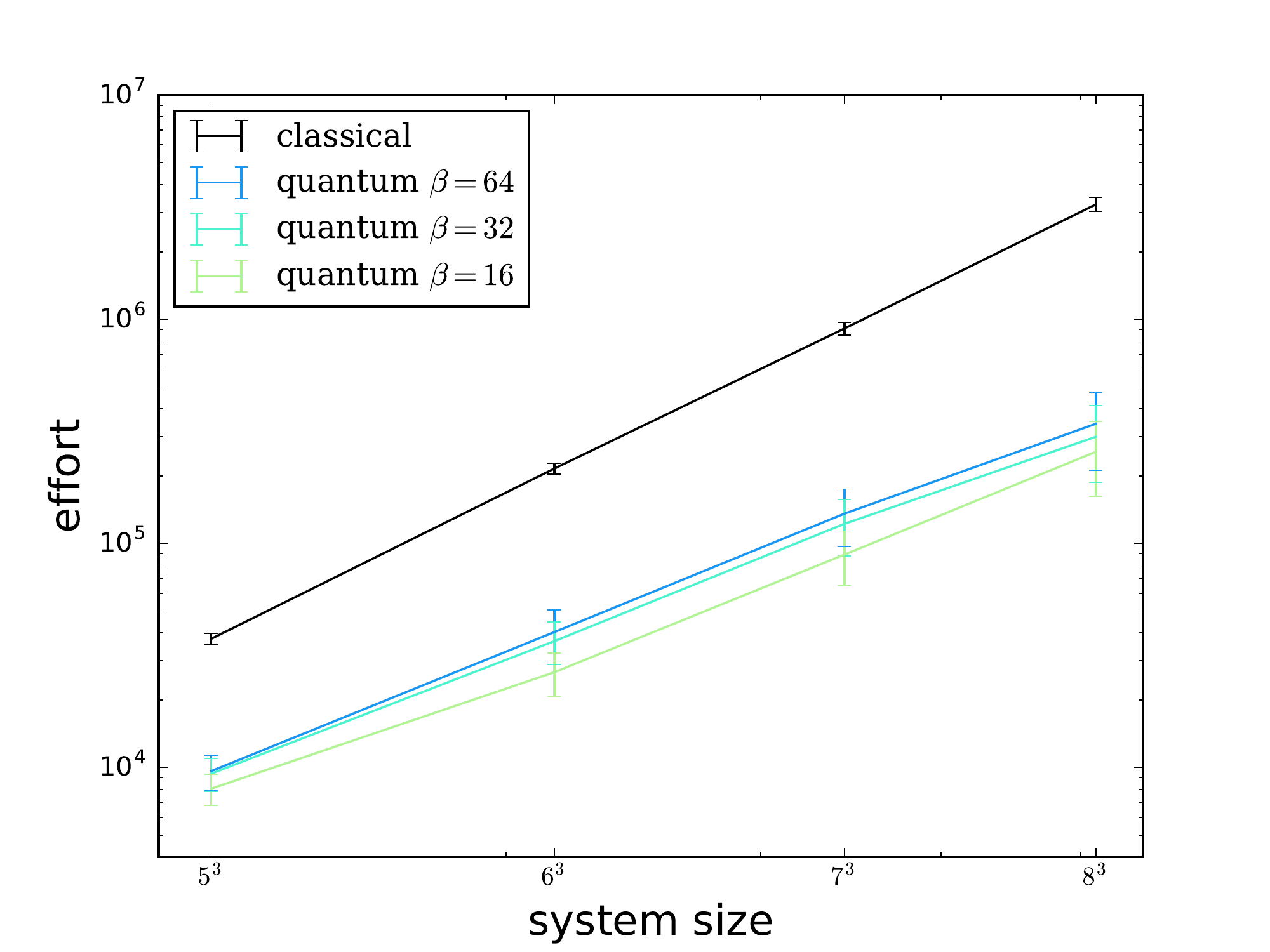}
 \caption{\label{fig:scaling}Scaling properties for QA and CA: Median effort to find the ground state with probability $90\%$. Since these are simulations, the performance of the real machine may differ by some constant factor. Still, a scaling advantage can be observed. For a confidence level of $95\%$ our regression gave a classical slope of $0.650 \pm 0.003$ and for the quantum case the slope is given by $0.51 \pm 0.06$ for $\beta = 16$. The slope of $\beta=32$ is given by $0.52 \pm 0.06$, and the slope of $\beta=64$ is given by $0.54 \pm 0.07$ both are calculated with the same confidence as for CA.}
\end{figure}

Results for repeated fast annealing are shown in Fig.~\ref{fig:scaling}. We observe a slight scaling advantage for SQA over CA. 


Apart from the proposed adaptive schedule, we performed a comparison of different nonlinear parameterizations of annealing schedules for QA, but found for optimized parameters no speedup compared to the linear schedule for any of them. The analysis for these and further simulations can be found in the supplementary material.

\paragraph{Conclusion}

For fair comparisons both CA and QA need to be optimized, otherwise wrong conclusions for the efficiency are drawn. To achieve this goal we proposed a heuristic nonlinear schedule and demonstrated that it is resistant against sub-optimally chosen initial values for the transverse field $\Gamma_0$ on random 3D Ising spin glasses.

A comparison between CA and QA for 3D Ising spin glasses gives similar results as in 2D. The existence of a finite temperature phase transition does not have a major influence on the relative performance between CA and QA. While QA rapidly finds a low energy local minimum, increasing the annealing time leads to CA finding lower energy states.
A similar conclusion is drawn in Ref.~\cite{zanca2015quantum}, where quantum speed-up is obtained in the random ferromagnetic Ising chain model, i.e.\ a system for which CA encounters no phase-transition at any finite temperature, while QA does.

Using many short runs, our simulations showed an advantage for the scaling of SQA over CA. Our simulations were restricted to relatively small system sizes, because the exact ground state had to be calculated beforehand. Surface effects of these small instances might skew the observed scaling behavior such that a verification of our results for larger instances should be performed.

Extending our an analysis to different problem classes will give further insight to where (S)QA gains an advantage over CA.

\paragraph{Acknowledgments}
We thank the spinglass server~\cite{sgserver} for their calculation of the ground state energy. We also thank Ilia Zintchenko for interesting discussions and the whiplash framework for easier management of measurement data. MT acknowledges hospitality of the Aspen Center for Physics, supported by NSF grant PHY-1066293.
This paper is based upon work supported in part by the Swiss National Science Foundation through NCCR QSIT and by ODNI, IARPA via MIT Lincoln Laboratory Air Force Contract No. FA8721-05-C-0002. 

\bibliography{biblio}

\end{document}


\title{Supplementary Material for: Optimizing schedules for Quantum Annealing}
\author{Daniel Herr}
\affiliation{Theoretische Physik, ETH Zurich, 8093 Zurich, Switzerland}
\affiliation{Quantum Condensed Matter Research Group, CEMS, RIKEN, Wako-shi 351-0198, Japan}
\author{Ethan Brown}
\affiliation{Theoretische Physik, ETH Zurich, 8093 Zurich, Switzerland}
\author{Bettina Heim}
\affiliation{Theoretische Physik, ETH Zurich, 8093 Zurich, Switzerland}
\author{Mario K\"onz}
\affiliation{Theoretische Physik, ETH Zurich, 8093 Zurich, Switzerland}
\author{Guglielmo Mazzola}
\affiliation{Theoretische Physik, ETH Zurich, 8093 Zurich, Switzerland}
\author{Matthias Troyer}
\affiliation{Theoretische Physik, ETH Zurich, 8093 Zurich, Switzerland}
\affiliation{Quantum Architectures and Computation Group, Microsoft Research,
Redmond, WA 98052, USA}

\begin{abstract}
This document contains additional plots proving the validity of the results in the main paper and plots showing intermediate measurements used for the adaptive schedule. Furthermore different parametrization for nonlinear annealing schedules are tested.
\end{abstract}
\maketitle

\subsection*{Methods used in the main paper}
In the following a more thorough description of algorithms used to simulate both Classical annealing (CA) and Quantum annealing (QA) is given.
\subsubsection*{Classical Annealing}
In CA~\cite{kirkpatrick1983} the system is first initialized to a random configuration at a high temperature and then gradually cooled.  This allows the system to escape from local minima and to relax towards lower energy configurations~\cite{sa_and_sa_conv_cond}.
During each step the Metropolis-Monte-Carlo algorithm~\cite{Metropolis} allows re-equilibration. Each step of this algorithm proposes a change to the current configuration. The change is always accepted if favorable, whereas if the cost increases it will only be accepted by a probability dependent on the temperature of the system.

\subsubsection*{Quantum Annealing}
During quantum annealing (QA)~\cite{Ray1989,Finnila1994,Kadowaki1998,idea_of_qa,qareview} the adiabatic theorem enables the system to evolve from a trivial ground state to the solution of the complex problem $\mathcal{H}_P$, if the change in the Hamiltonian is sufficiently slow.
For this one can state a new system Hamiltonian
\begin{equation}
    \mathcal{H}=s \mathcal{H}_P +  \left(1-s\right) \mathcal{H}_D
    \label{eq:generalhamil}
\end{equation}
that introduces the control parameter $s \in \left[0 ,1\right]$ which enables a transition from an initial system $\mathcal{H}_D$ whose ground state can be easily obtained to the problem Hamiltonian.
In contrast to CA, the system is held at a constant inverse temperature $\beta=1/T$ while the control parameter is slowly increased to 1. Thus, the QA schedule takes the system from a preparable state to a state where quantum fluctuation are suppressed and the system remains frozen in its configuration. The problem Hamiltonian encodes the classical problem that is about to be solved. In QA classical values of $s_i$ are replaced by the Pauli spin-z operators $\sigma_i^z$. And the driver Hamiltonian $\mathcal{H}_D$ introduces randomness to the quantum and is given by
\begin{equation}
  \mathcal{H}_D = \Gamma_0 \sum_i \sigma^x_i.
\end{equation}
Here $\sigma_i^x$ is the Pauli spin-x operator for the $i$th spin and enable transitions between $\uparrow$ and $\downarrow$ for a nonzero transverse field $\Gamma_0$.

To simulate QA we used a discrete time SQA algorithm. We use the path integral Monte Carlo (PIMC) method to map the 3 dimensional system into $3+1$ dimensions by the introduction of the imaginary time dimension~\cite{suzuki_orig}.
This extra dimension is discretized with $M$ Trotter slices, which are copies of the classical Ising system coupled to each other by a value dependent on $\Gamma$ and $\beta$.
As the transverse field decreases the coupling between the Trotter slices gets stronger. In the classical limit $\Gamma = 0$ the quantum fluctuation are suppressed and the system remains frozen in its configuration.
The simulation using a finite number of Trotter slices $M$ is called discrete time SQA (DT-SQA) and comes with a discretization error of $O(\beta^3/M^2)$.
The physical limit corresponds to the continuous time limit $M\rightarrow \infty$. To compare the computational effort between QA and CA the number of Monte Carlo sweeps of SQA and CA are compared. In SQA a single update consists of a Swendsen-Wang~\cite{swendsenwang} cluster move in imaginary time direction only opposed to the single spin flip CA performs. This has proven to be a reliable comparison for the individual runtime~\cite{Heimb15}.

Since recent work by some of the authors indicates that using a PIMC technique with open boundary conditions in imaginary time is the best possible SQA algorithm and may better reproduce the scaling with system's size of a coherent quantum annealer~\cite{mazzola2017quantum,Guglielmo}, the simulations were run accordingly. The results serve as a best case estimate for the performance of a physical quantum annealer, e.g.\ the D-Wave 2, since the coherence length in such devices may or may not extend across the entire system. Furthermore all simulations were run in the physical limit where the number of Trotter slices is required to be high enough to ensure convergence. This leads to $M=1024$.

\subsection*{Observables in PIMC}
To evaluate the expectation value of the QA counterpart of the specific heat $C_q$, we need to compute the value of $\left<\sigma_x\right>$. A derivation on how to calculate this observable using the PIMC method can be found in~\cite{Krzakala08} and will be reproduced in the following.\\
Generally an observable can be evaluated as:
\begin{equation}
    \left<O\right> = \frac{1}{Z} \text{tr}\left( O e^{-\beta \mathcal{H}}\right)
\end{equation}
Thus the expectation value of $\left<\sigma_x\right>$ will evaluate to the following:
\begin{equation*}
    \left<\sigma_i\right>=\frac{\sum_{\sigma_i} \bra{\sigma_i} \sigma_i^x e^{-\beta\mathcal{H}}\ket{\sigma_i}}{\sum_{\sigma_i} \bra{\sigma_i} e^{-\beta \mathcal{H}} \ket{\sigma_i}}
\end{equation*}
Introducing the Trotter decomposition $e^{\beta\mathcal{H}} = {\left( e^{\frac{\beta}{M} \mathcal{H}}\right)}^M$ with $\tau = \frac{\beta}{M}$ and adding identity operators will give
\begin{equation}
    \left<\sigma_i\right>\Big|_k = \frac{\sum_{\sigma_i} \bra{\sigma_i} \sigma_i^x e^{-\beta\mathcal{H}}\ket{\sigma_i}}{\sum_{\sigma_i} \bra{\sigma_i} e^{-\beta \mathcal{H}} \ket{\sigma_i}}.
\end{equation}
But there is freedom to choose for which Trotter slice to evaluate $\sigma_i^x$. So one can also take the average of all these choices:
\begin{equation}
    \left<\sigma_i\right>= \frac{1}{M} \sum_{k=1}^{M} \left<\sigma_i\right>\Big|_k
\end{equation}
Now with the introduction of a small error of order $O(\tau^2)$ one can split $e^{-\tau (\mathcal{H}_p + \Gamma\sum_i\sigma_i^x)}= e^{-\tau \mathcal{H}_p} e^{-\tau \Gamma \sum_i \sigma_i^x}$ and then find for the expectation value
\begin{equation}
    \left<\sigma_i^x\right> = \frac{1}{M}\sum_{k=1}^{M} \left( \frac{\bra{\sigma_i^{k+1}} \sigma_i^x e^{-\tau \Gamma \sum_i \sigma_i^x }\ket{\sigma_i^k}}{\bra{\sigma_i^{k+1}} e^{-\tau \Gamma \sum_i \sigma_i}\ket{\sigma_i^k}} \right)
\end{equation}
Using
$\bra{\uparrow} e^{a\sigma^x}\ket{\uparrow} = \cosh(a)$
and
$\bra{\uparrow} e^{a\sigma^x}\ket{\downarrow} = \sinh(a)$
one can obtain the expectation value of the Pauli spin x operator:
\begin{equation}
    \left<\sigma_i^x\right>= \frac{1}{M} \sum_{k=1}^{M} {\tanh\left(-\tau \Gamma \right)}^{s_i^k s_i^{k+1}}
\end{equation}

\begin{figure}
  \includegraphics[width=\linewidth]{sigmax.pdf}
  \caption{\label{fig:sigmax} The expectation value of $\sigma_x$ operator depending on $\Gamma$}
\end{figure}

In Figure~\ref{fig:sigmax} the expectation value of $\sigma_x$ is plotted for different values of the transverse field.
For the more specific schedule relying on the Hamiltonian
\begin{equation*}
    \mathcal{H} = \mathcal{H}_p + \Gamma(t) \sum_{i=0}^N \sigma_x^i
\end{equation*}
one can get an expectation value for $C_q$ by the following formula:
\begin{equation*}
  C_q = \frac{1}{\beta} \frac{d^2}{d\Gamma^2} \log(Z)= \beta\Gamma\left(1 - \left<\sigma_x\right>^2\right)
\end{equation*}
Finally one should note that any constant prefactor in the measurement of $C_q$ will cancel further on when $\lambda$ is determined by fixing the number of MCS.\@

\subsection*{Schedules}
\begin{figure}
  \includegraphics[width=\linewidth]{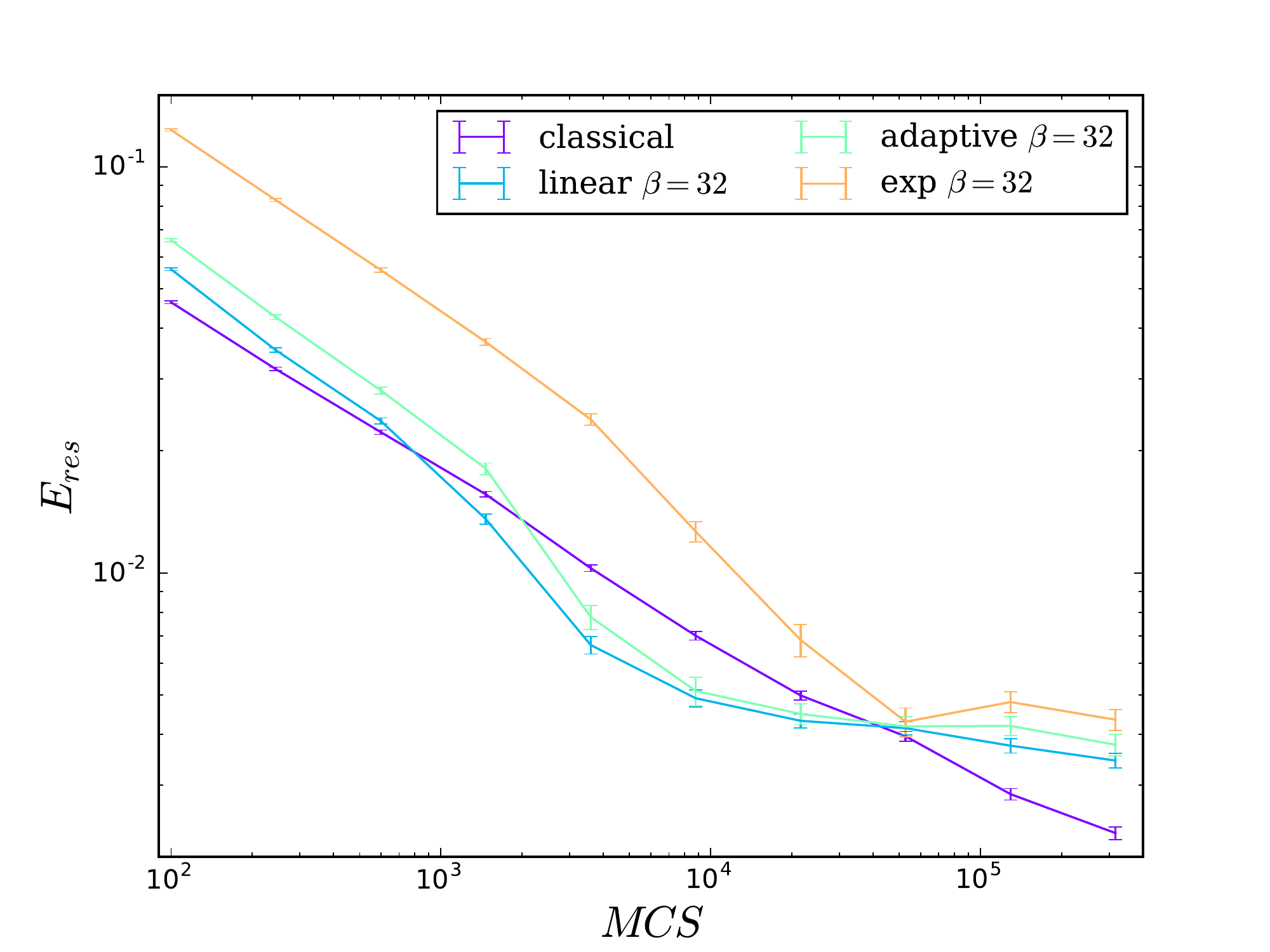}
  \caption{\label{fig:nonlin_perf}Performance of the different schedules for optimized start and end parameters}
\end{figure}

\begin{figure}
  \includegraphics[width=\linewidth]{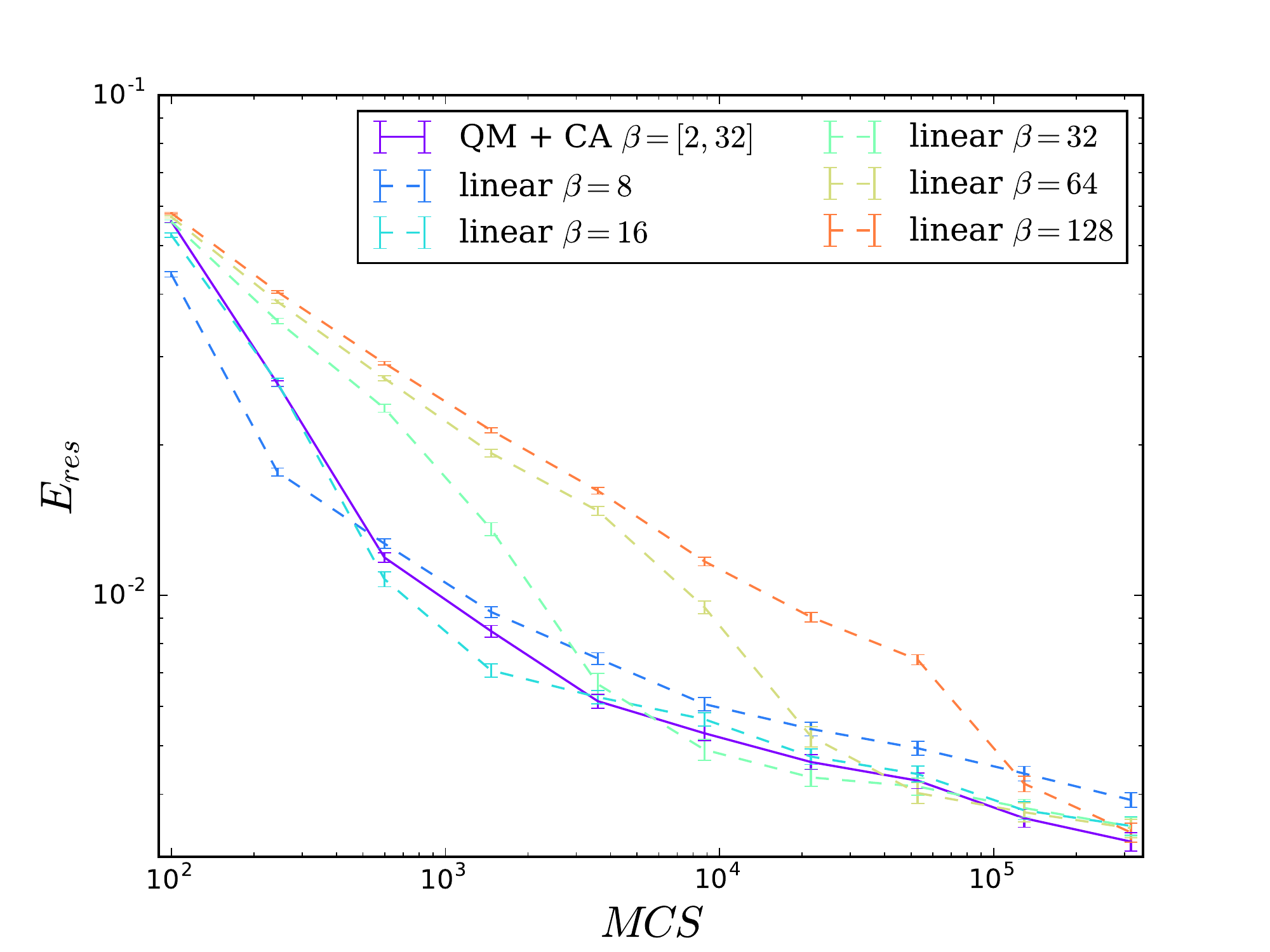}
  \caption{\label{fig:BG_sched}Median residual energy for a schedule with linear change in both $\beta$ and $\Gamma$}
\end{figure}

For optimized parameters an exponential schedule and a mixture of both classical and quantum annealing, where the transverse field was linearly decreased while the inverse temperature was linearly increased, were tested.
The performance of these schedules is plotted in Figure~\ref{fig:nonlin_perf}. Even the hybrid between CA and QA did not show a considerable performance increase, as can be seen in Figure~\ref{fig:BG_sched}.
The additional parameters that need to be optimized only seem to result in marginal changes. The residual energy seems behave similarly to SQA runs with fixed high temperature at few MCS, while still obtaining a slightly lower state which low temperature runs possess for large numbers of MCS.\@
\begin{figure}
  \includegraphics[width=\linewidth]{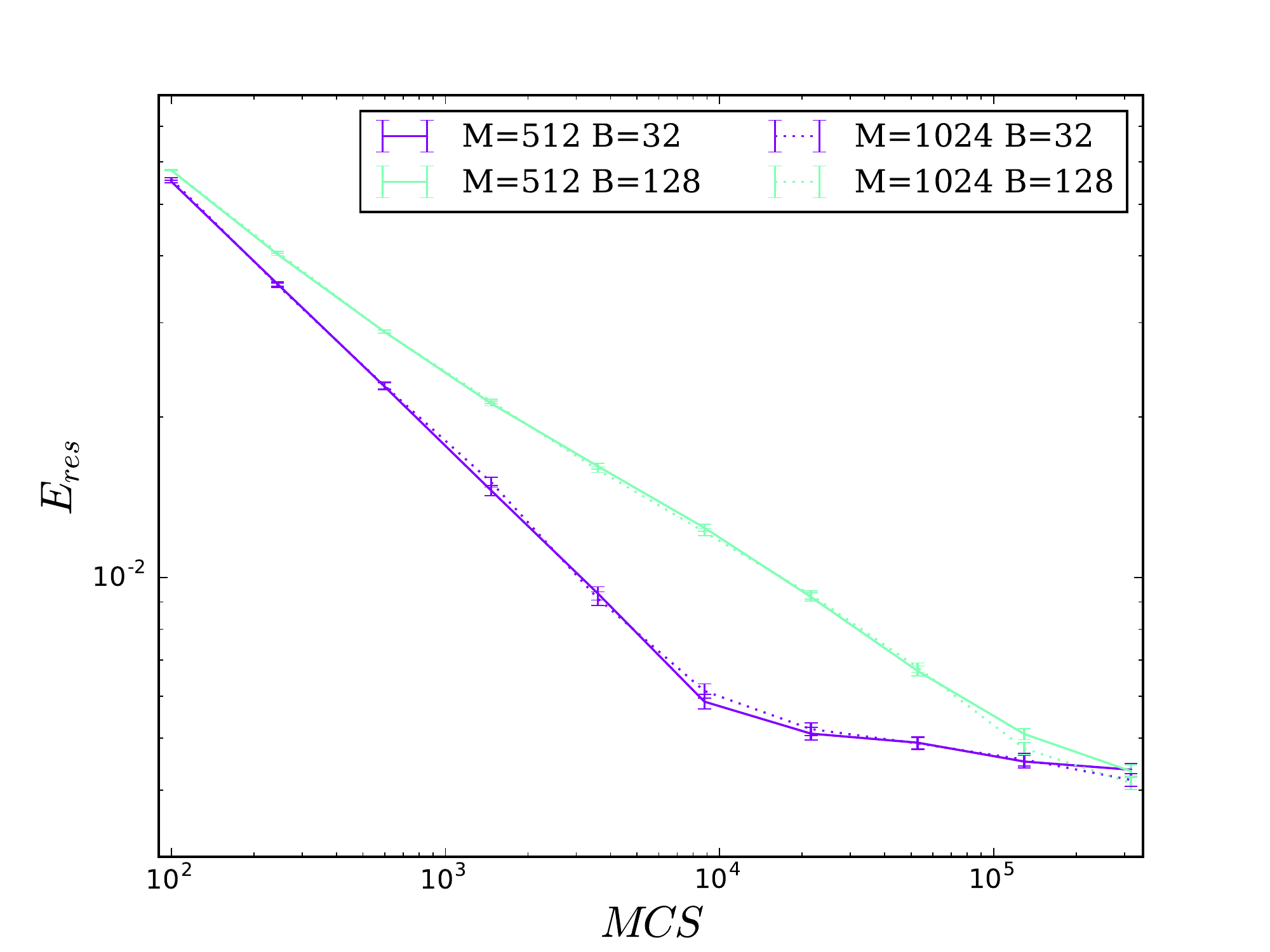}
  \caption{\label{fig:convergence}Change in number of trotter slice to check if DT-SQA already converged}
\end{figure}

\subsection*{Notes on the adaptive schedule}
The adaptive schedule presented in this paper is derived using the system Hamiltonian given in Eq.~\ref{eq:generalhamil}. Yet, a more common annealing procedure~\cite{Heimb15,santoro2} is when the problem Hamiltonian is kept at constant strength while the transverse field is decreased from $\Gamma_0$ to $0$. The simulations performed in this paper use this more specific schedule.

\subsection*{Validity of Results}
In Figure~\ref{fig:convergence} one can see that at 1024 Trotter slices a change in discretization has almost no influence on the behavior of QA.\@ Thus one can conclude that the algorithm is sufficiently converged.

\subsection*{Ferromagnet}
\begin{figure}
  \includegraphics[width=\linewidth]{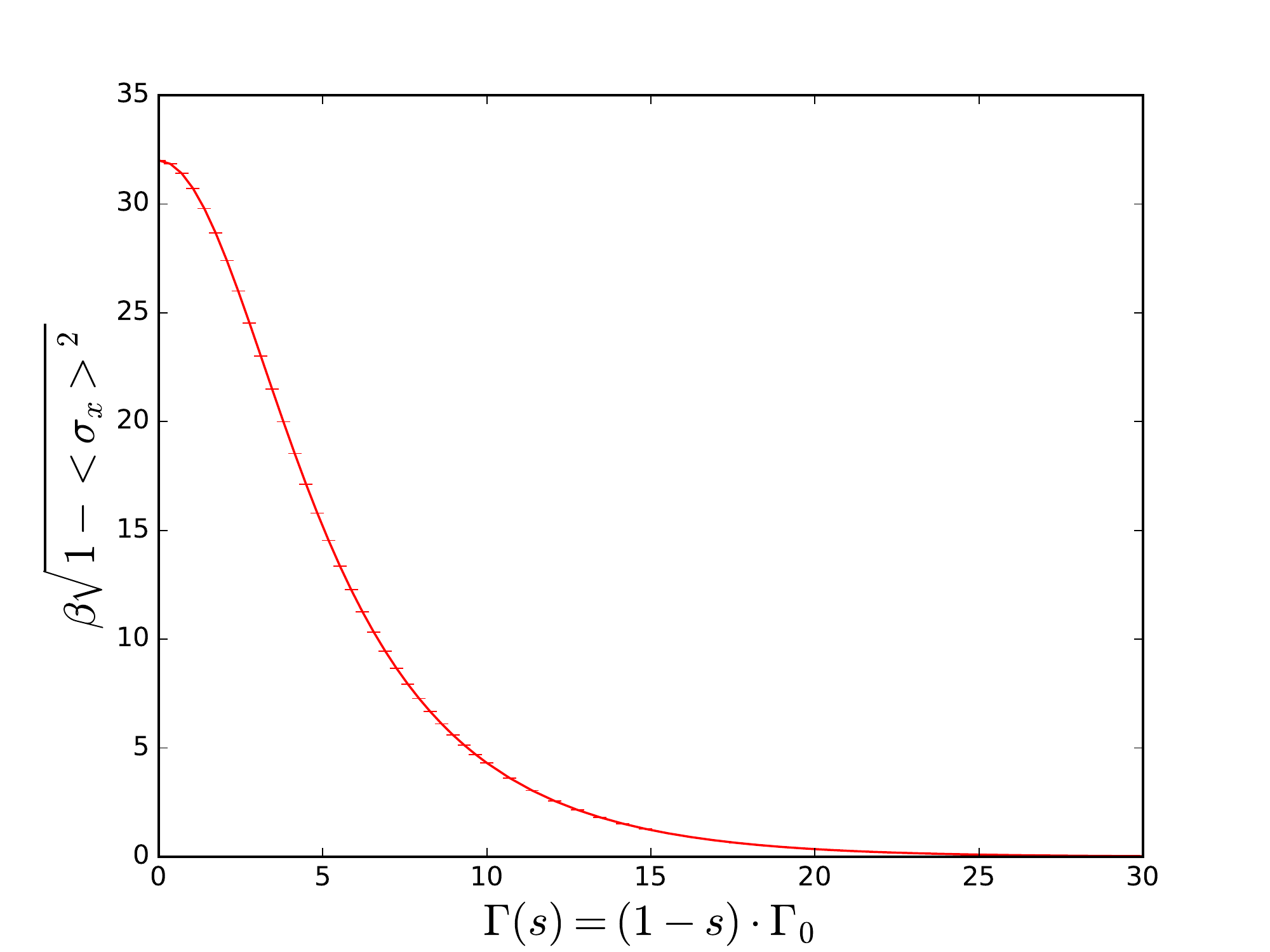}
  \caption{\label{fig:sched_ferro}Measured indicator for the annealing speed of a 3D Ferromagnet with $\beta = 32$}
\end{figure}

\begin{figure}
  \includegraphics[width=\linewidth]{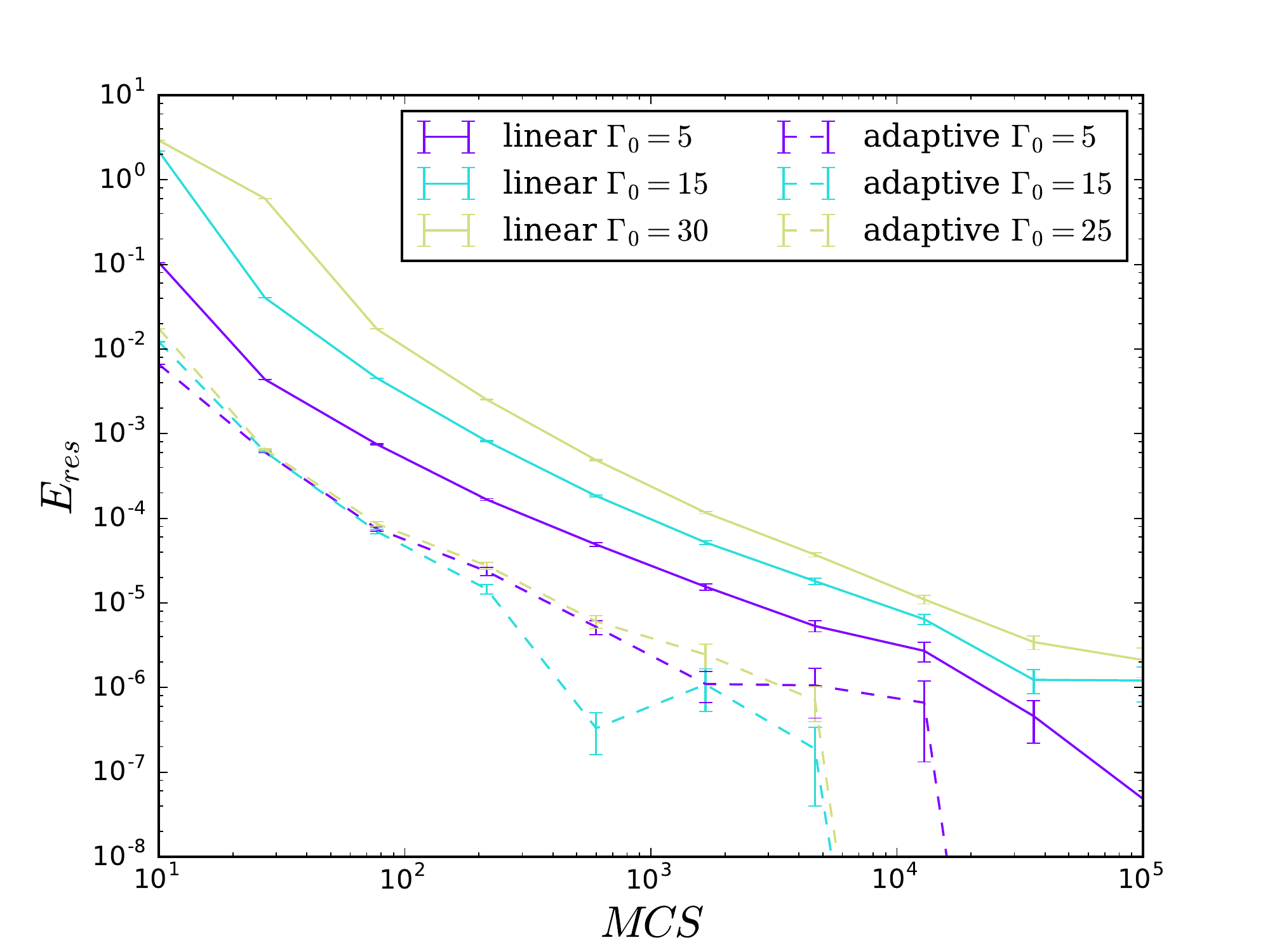}
  \caption{\label{fig:Ferro_perf}Performance comparison between the adaptive and linear for a 3D Ferromagnet}
\end{figure}

Additionally the same analysis was conducted for a 3D Ferromagnet whose degenerate ground state was lifted by a local field in $\sigma_z$ direction. The parameter for the annealing speed is plotted in Figure~\ref{fig:sched_ferro} and in Figure~\ref{fig:Ferro_perf} for different unoptimized annealing parameters. Again one can see that the adaptive schedule is better than the linear schedule for unoptimized parameters. But we noticed that optimized values, can further improve the performance such that a well optimized starting value beats the unoptimized adaptive schedule in the case of a Ferromagnet. Still this schedule becomes more advantageous the farther from optimized start values the annealing run is.


\subsection*{Comparison to 2D}
\begin{figure}
  \includegraphics[width=\linewidth]{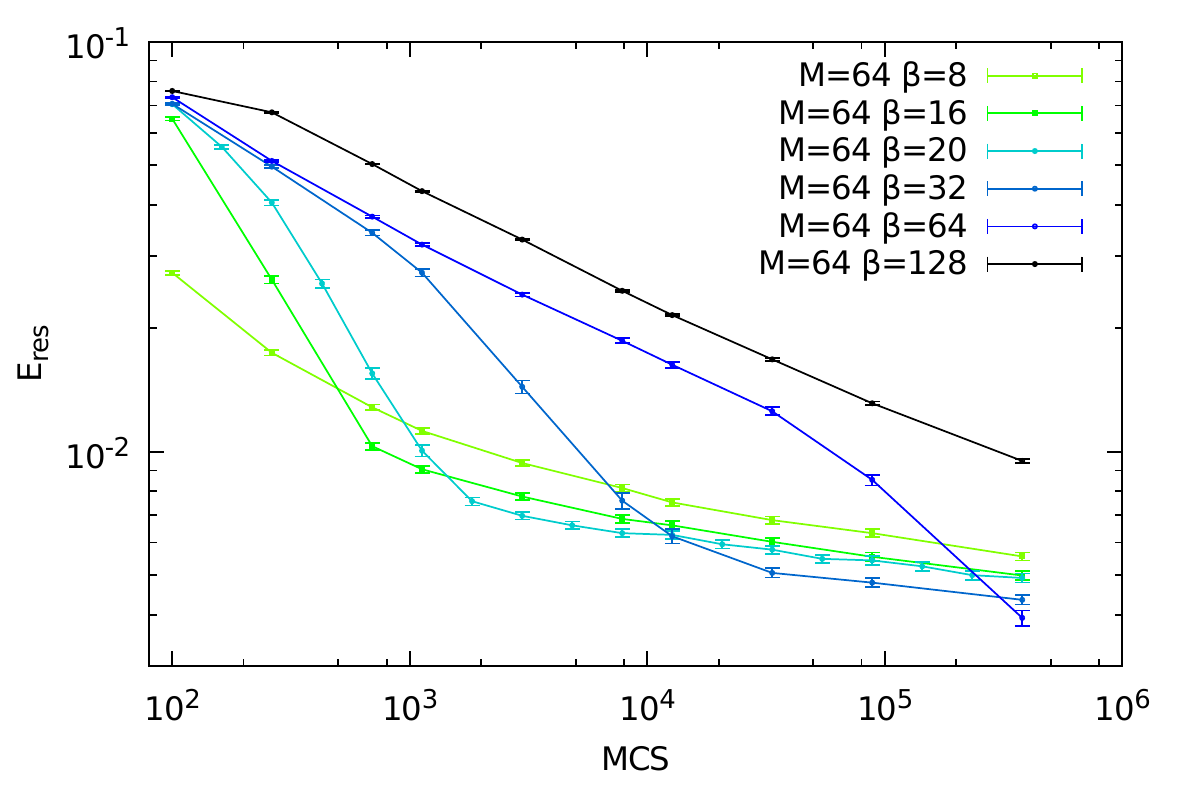}
  \caption{\label{fig:bettina_B}temperature behavior of the DT-SQA algorithm without the requirement of convergence to the continuous limit}
\end{figure}

Simulations not in the converged limit show the exact same behavior as for 2D Ising spin glasses, cf. Figure~\ref{fig:bettina_B} and the paper on 2D Ising spin glasses~\cite{Heimb15}. The same can be observed for the converged case, that is presented in the main paper.

\bibliography{biblio}